\documentstyle[preprint,eqsecnum,aps,prd]{revtex}
\begin{document}
\draft \preprint{DAMTP 95-32} \title{Approximate and Exact Consistency
  of Histories.} \author{J. N.  McElwaine\thanks{E-mail:
  J.N.McElwaine@damtp.cambridge.ac.uk}} \address{Department of Applied
  Mathematics and Theoretical Physics,\\ University of Cambridge,\\
  Silver Street, Cambridge CB3 9EW, U.K.}

\date{To appear Phys. Rev. A, 1st March 1996} \maketitle
\begin{abstract}
  The consistent histories formalism is discussed using path-projected
  states. These are used to analyse various criteria for approximate
  consistency. The connection between the Dowker-Halliwell criterion
  and sphere packing problems is shown and used to prove several new
  bounds on the violation of probability sum rules. The quantum Zeno
  effect is also analysed within the consistent histories formalism
  and used to demonstrate some of the difficulties involved in
  discussing approximate consistency. The complications associated
  with null histories and infinite sets are briefly discussed.
\end{abstract}
 \narrowtext
\section{Introduction}

The Copenhagen interpretation of Quantum mechanics occupies a very
unusual place among physical theories: it contains classical mechanics
as a limiting case, yet at the same time it requires this limiting
case for its own interpretation \cite{LL:QM,Espagnat}. This problem is
particularly acute in quantum cosmology, since it is highly unlikely
that any systems obeying classical mechanics existed in the early
universe.

The consistent histories approach to quantum mechanics is an attempt
to remove the ambiguities and difficulties inherent in the Copenhagen
interpretation. The basic objects are sequences of events or {\em
histories\/}. A set of histories must include all possibilities and
must be {\em consistent\/}. The individual histories can then be
considered physical possibilities with definite probabilities, and
they obey the ordinary rules of probability and logical inference. The
predictions of the consistent histories formalism are identical to the
predictions of standard quantum mechanics where laboratory experiments
are concerned, but they take place within a more general theory.

Much work has been done on trying to understand the emergence of
classical phenomena within the consistent histories approach
\cite{GM:Hartle:classical,Caldeira:Leggett,Joos:Zeh,%
Dowker:Halliwell,Pohle,Paz:brownian:motion,%
Zurek:preferred:observables,Anastopoulos:Halliwell,%
Tegmark:Shapiro,Paz:Zurek:Classicality,Hartle:spacetime,%
Zurek:transition}. These studies consider closed quantum systems in
which the degrees of freedom are split between an unobserved
environment and distinguished degrees of freedom such as the position
of the centre of mass.

In these and other realistic models it is often hard to find
physically interesting, {\em exactly\/} consistent sets, so most
examples studied are only approximately consistent. These models do
show, however, that histories consisting of projections onto the
distinguished degrees of freedom at discrete times are {\em
approximately\/} consistent.  This work is necessary for explaining
the emergence of classical phenomena but is incomplete. The
implications of different definitions differences of {\em
approximate\/} consistency have received little research: the subject
is more complicated than has sometimes been realised.  A quantitative
analysis of the quantum Zeno paradox demonstrates some of the
problems. A deeper problem is explaining why quasi-classical sets of
histories occur as opposed to any of the infinite number of
consistent, non-classical sets. Until these problems are understood
the program is incomplete.

In this paper I examine two different criteria approaches to
approximate consistency and analyse two frequently used criteria. I
show a simple relation with sphere-packing problems and use this to
provide a new bound on probability changes under coarse-grainings.

\subsection{Consistent Histories Formalism}
The most basic objects in the consistent histories formulation are
projection operators, representing particular states of affairs
existing at particular times \cite{Griffiths:1}.  These are combined
into time-ordered strings which are the elementary events, or {\em
histories\/}, in the probability sample space $S$. A set of projective
decompositions of the identity $\{\sigma_n,\ldots,\sigma_1\}$ and
times $\{t_n,\ldots,t_1\}$ define a set of {\em class-operators\/} $S
= \{C_\alpha\}$,
\begin{eqnarray}\nonumber 
  C_\alpha &=& U(-t_n) P^n_{\alpha_n} U(t_{n}-t_{n-1})
  P^{n-1}_{\alpha_{n-1}} \ldots \\ &&U(t_2-t_1)P^1_{\alpha_1}U(t_1),
  \label{type1}
\end{eqnarray}
where $P^j_{\alpha_j} \in \sigma_j$ and $U(t)$ is the time evolution
operator. The explicit time dependence of this set can be removed by
defining new sets of projectors $\sigma_k' =
U(-t_k)\sigma^k_{\alpha_k}U(t_k$).

More general sets of class-operators can be created by {\em
  coarse-graining. \/}$S^* = \{C^*_\beta\}$ is a coarse-graining of
  $S$ if $ C^*_\beta = \sum_{\alpha \in \overline{\alpha}_\beta}
  C_\alpha$, where $\{\overline{\alpha}_\beta\}$ is a partition of S.
  Omn\`es defines sets of histories without any coarse-graining as
  Type I, and those which have been coarse-grained but where the
  class-operators are still strings of projectors as Type II
  \cite{Omnes:1}. I shall follow Isham\cite{Isham:homog} and call
  these {\em homogenous}.

Gell-Mann and Hartle consider completely general coarse-grainings, and
also ones which they call branch-dependent\cite{GM:Hartle:classical}.
These are a restriction of Type II histories to those in which earlier
projections are independent of later ones.
 
The type of histories used makes very little difference to discussions
of approximate consistency. I will therefore follow Gell-Mann and
Hartle and use completely general class-operators, though on occasion
I shall state stronger results which hold for homogenous
class-operators.

Probabilities are defined by the formula
\begin{equation}
  P(\alpha) = D_{\alpha\alpha},
\end{equation}
where $D_{\alpha\beta}$ is the decoherence matrix
\begin{equation}\label{dm}
  D_{\alpha\beta} = \mbox{Tr} (C_\alpha\rho {C_\beta}^\dagger),
  \label{dmdef}
\end{equation}
and where $\rho$ is the initial density
matrix\footnote{Generalisations exist that also have a final density
matrix\cite{Griffiths:1}.}. If no further conditions were imposed
these probabilities could contradict ordinary quantum mechanics: they
would be inconsistent.  A necessary and sufficient requirement for
consistency is that the probability of a collection of histories
should not change under {\em any\/} coarse-graining
\cite{CEPI:GMH}\footnote {Griffiths and Omn\`es only consider more
restricted sets of histories, which results in a weaker condition
\cite{Griffiths:1,Omnes:1}.}. This condition can be expressed
\begin{equation}
  \mbox{Re}(D_{\alpha\beta}) = 0, \quad \forall \alpha \neq \beta,
  \label{weakcon}
\end{equation}
which Gell-Mann and Hartle call {\em weak consistency\/}. A stronger
condition,
\begin{equation}
  D_{\alpha\beta} = 0, \quad \forall \alpha \neq \beta,
  \label{mediumcon}
\end{equation}
is often used in the literature for simplicity\footnote{Gell-Mann and
  Hartle call this medium consistency, and also define two even
  stronger conditions \cite{GM:Hartle:classical}.}. I shall restrict
  my discussion to the weak condition (\ref{weakcon}), as it is
  necessary physically and any consequent results will certainly hold
  if a stronger condition is satisfied.

\subsection{Path-Projected States}

A simple way of regarding a set of histories is as a set of
path-projected states or {\em history states\/}\footnote{This approach
loses its advantages if a final density matrix is present.}. For a
pure initial density matrix $\rho = |\psi\rangle\langle\psi|$ these
states are defined by
\begin{equation}  \label{PPS}
  {\bf u_\alpha} = C_\alpha|\psi\rangle\, \in \, {\bf H_1},
\end{equation}
where $\mbox{dim}({\bf H_1}) = d$. For a mixed density matrix,
\begin{equation}
  \rho = \sum_{i=1}^n p_i|\psi_i\rangle_1\langle\psi_i|_1 \qquad
  |\psi_i\rangle_1 \, \in \, {\bf H}_1,
\end{equation}
history states can be defined by regarding $\rho$ as a reduced density
matrix of a pure state in a larger Hilbert space ${\bf H}_1\otimes{\bf
H}_2$, where ${\bf H}_2$ is of dimension $\mbox{rank}(\rho) = n$
(possibly infinite), with orthonormal basis $|i\rangle_2$. All
operators $A_1$ on ${\bf H}_1$, can be extended to operators on ${\bf
H}_1\otimes{\bf H}_2$ by defining $A = A_1\otimes{\bf 1}_2$.  The
state in the larger space is
\begin{equation}
  |\psi\rangle = \sum_{i=1}^n \sqrt{p_i} |\psi_i\rangle_1 \otimes
  |i\rangle_2 \qquad |\psi\rangle \, \in \, {\bf H}_1\otimes{\bf H}_2,
\end{equation}
and the history states are again given by equation (\ref{PPS}); but
now they are vectors in an $N = nd$ dimensional Hilbert space.

The decoherence matrix (\ref{dm}) is
\begin{equation}
  D_{\alpha\beta} = \mbox{Tr}({\bf u_\alpha}{\bf u_\beta^\dagger}) =
  {\bf u_\beta^\dagger}{\bf u_\alpha},
\end{equation}
so the probability of the history $\alpha$ occurring is $||{\bf
  u_\alpha}||^2$. The consistency equations (\ref{weakcon}) are
\begin{equation} \label{weakvectorcon}
  \mbox{Re}({{\bf u}_\alpha}^\dagger{\bf u}_\beta) = 0 \quad
  \forall\,\alpha \neq \beta.
\end{equation}

A complex Hilbert space of dimension $N$ is isomorphic to the real
Euclidean space ${\bf R}^{2N}$. The consistency condition
(\ref{weakvectorcon}) takes on an even simpler form when the history
states are regarded as vectors in the real Hilbert space. I define the
{\em real history states \/}
\begin{equation} \label{realsplit}
  {\bf v}_{\alpha} = \mbox{Re}({\bf u}_{\alpha}) \oplus \mbox{Im}({\bf
    u}_{\alpha}) \quad \in {\bf R}^{2N},
\end{equation} 
and then the consistency condition (\ref{weakvectorcon}) is that the
set of real history states, $\{{\bf v}_{\alpha}\}$, is orthogonal,
\begin{equation} \label{realweak}
  {\bf v}_\alpha^T{\bf v_\beta} = 0 \quad \forall \, \alpha \neq
\beta.
\end{equation}
The probabilities of history $\alpha$ is $||{\bf v}_\alpha||^2$.

For the rest of this paper I shall only consider pure initial states
since the results can easily be extended to the mixed case by using
the above methods.

\section{Approximate Consistency}
In realistic examples it is often difficult to find physically
interesting, {\em exactly\/} consistent sets. This rarity impacts upon
the use of consistent histories in studies of dust particles or
oscillators coupled to environments
\cite{GM:Hartle:classical,Caldeira:Leggett,Joos:Zeh,%
Dowker:Halliwell,Pohle,Paz:brownian:motion,%
Zurek:preferred:observables,Anastopoulos:Halliwell,%
Tegmark:Shapiro,Paz:Zurek:Classicality,Hartle:spacetime,%
Zurek:transition}.  Frequently in these studies, the off-diagonal
terms in the decoherence matrix decay exponentially with the time
between projections, but their real parts are never exactly zero, so
the histories are only approximately consistent. Therefore if the
histories are coarse-grained, the probabilities for macroscopic events
will vary very slightly depending on the exact choice of histories in
the set.  Because the probabilities can be measured experimentally,
they should be unambiguously predicted --- at least to within
experimental precision.

In his seminal work Griffiths states that ``violations of [the
consistency criterion (\ref{weakcon})] should be so small that
physical interpretations based on the weights [probabilities] remain
essentially unchanged if the latter are shifted by amounts comparable
with the former''\cite[sec. 6.2]{Griffiths:1}. Omn\`es
\cite{Omnes:1,Omnes:interpretation,Omnes:2,Omnes:3,Omnes:4,%
Omnes:5,Omnes:truth}, and Gell-Mann and Hartle
\cite{GM:Hartle:classical,CEPI:GMH,Hartle:QM:Cosmology} make the same
point. The amount by which the probabilities change under
coarse-graining is the extent to which they are ambiguous. I shall
define the the largest such change in a set to be the {\em maximum
probability violation\/} or MPV.

Dowker and Kent\cite{Dowker:Kent:approach,Dowker:Kent:Properties}
argue that more is needed.  Why should {\em approximately\/}
consistent sets be used?  They suggest that ``near'' a generic
approximately consistent set there will be an exactly consistent one.
``Near'' means that the two sets describe the same physical events to
order $\epsilon$; the relative probabilities and the projectors must
be the same to order $\epsilon$. In the paper I investigate which
criteria will guarantee this, and show that some of the commonly used
ones are not sufficient.

\subsection{Probability Violation}

The MPV can be defined equivalently in terms of the decoherence
matrix:
\begin{eqnarray}\label{MPV}
  \mbox{MPV}(D) & = &\max_{\overline \alpha} \left| P(\overline
  \alpha) - \sum_{\alpha \in \overline \alpha} P(\alpha) \right|, \\ &
  = & \max_{\overline \alpha} \left| \sum_{\alpha,\beta \in \overline
  \alpha} D_{\alpha\beta} - \sum_{\alpha \in \overline \alpha}
  D_{\alpha\alpha} \right|, \\ & = & \max_{\overline \alpha} \left |
  \sum_{\alpha \neq \beta \in \overline \alpha}D_{\alpha\beta}
  \right|.
\end{eqnarray} 
The maximum is taken over all possible coarse-grainings $\overline
\alpha$.  For large sets of histories this is difficult to calculate
as the number of possible coarse-grainings is $O(2^n)$.  A simple
criterion that if satisfied to some order $\epsilon(\delta)$ would
ensure that the MPV were less than $\delta$ would be preferable here.

This is not a trivial problem. The frequently used criterion
\cite{GM:Hartle:classical,Omnes:5}
\begin{equation}
\label{badcriterion}
  |D_{\alpha\beta}| \leq \epsilon(\delta), \quad \forall \alpha \neq
\beta
\end{equation}
is not sufficient for any $\epsilon(\delta) > 0 $. Theorem
(\ref{proveuseless}) shows that for any $\epsilon(\delta) > 0 $ there
are finite sets of histories satisfying (\ref{badcriterion}) with an
arbitrarily large MPV. The example used in the proof also shows some
of the complications that arise when discussing infinite sets of
histories. All sets of histories in the rest of the paper will be
assumed to be finite unless otherwise stated.

A simple bound\footnote{When the class-operators are homogenous the
  bound can be improved to $M(D_{\alpha\beta}) \leq 1/2 \sum_{\alpha
  \neq \beta }|\mbox{Re}(D_{\alpha\beta})|$, since $\sum_{\alpha \neq
  \beta}D_{\alpha\beta} = 0$.} for the MPV is
\begin{equation} \label{modupbound}
  \mbox{MPV}(D) \leq \sum_{\alpha \neq \beta }|\mbox{Re}
  (D_{\alpha\beta})|.
\end{equation} 
This leads to the criterion for the individual elements
\begin{equation} \label{naivecriterion}
  |\mbox{Re}(D_{\alpha\beta})| \leq \frac{\delta}{n(n-1)}, \quad
  \forall \alpha \neq \beta,
\end{equation}
where $n$ is the number of histories.  Equation (\ref{naivecriterion})
ensures that the MPV is less than $\delta$, although the condition
will generally be much stronger than necessary. It would be preferable
however, to have a criterion that only depended on the Hilbert space
and not on the particular set of histories.

\section{The Dowker-Halliwell Criterion}
Dowker and Halliwell discussed approximate consistency in their paper
\cite{Dowker:Halliwell}, in which they introduced a new
criterion\footnote{ I have replaced Dowker-Halliwell's $<$ with $\leq$
to avoid problems with histories of zero probability.}
\begin{eqnarray}\label{DHC}
  \left|\mbox{Re}(D_{\alpha\beta})\right| & \leq & \epsilon \,
    (D_{\alpha\alpha}D_{\beta\beta})^{1/2}, \quad \forall\,
    \alpha\neq\beta,
\end{eqnarray}
which I shall call the {\em Dowker-Halliwell criterion\/} or DHC.
Using the central limit theorem and assuming that the off-diagonal
elements are independently distributed, Dowker and Halliwell
demonstrate that (\ref{DHC}) implies
\begin{eqnarray}\label{DHsumlaw}
  \left|\sum_{\alpha \neq \beta \in \overline\alpha}
    D_{\alpha\beta}\right| & \leq & \epsilon \sum_{\alpha \in
    \overline\alpha}D_{\alpha\alpha},
\end{eqnarray}
for most coarse-grainings $\overline\alpha$. This is a natural
generalisation of (\ref{weakcon}) to saying that the probability sum
rules are satisfied to {\em relative\/} order $\epsilon$. For
homogenous histories this is a similar but stronger condition than
requiring that the MPV (\ref{MPV}) is less than $\epsilon$, since
\begin{eqnarray}
  \sum_{\alpha \in \overline\alpha}D_{\alpha\alpha} &\leq & \sum
  D_{\alpha\alpha} = 1.
\end{eqnarray}
But for general class-operators $\sum D_{\alpha\alpha}$ is unbounded
and (\ref{DHsumlaw}) must either be modified or supplemented by a
condition such as
\begin{eqnarray}
  \left|\sum D_{\alpha\alpha}-1\right| & \leq & \epsilon.
\end{eqnarray}
This is only a very small change and for approximately consistent sets
is not significant.

For the sake of completeness, I shall occasionally mention a similar
criterion which I shall call the {\em medium DHC\/}
\begin{eqnarray}
  \label{MDHC}
  \left|D_{\alpha\beta}\right| & \leq & \epsilon \,
    (D_{\alpha\alpha}D_{\beta\beta})^{1/2}, \quad \forall\,
    \alpha\neq\beta.
\end{eqnarray}

As Dowker and Halliwell point out, \cite{Halliwellconversation}, the
off-diagonal terms are often not well modelled as independent random
variables. Indeed even when this assumption is valid, the MPV will
usually be much higher. By appropriately choosing $\epsilon$ as a
function of $\delta$, however, it is possible to eliminate these
problems, and to utilise the many other useful properties of the DHC.

\subsection{Geometrical Properties}
The Dowker-Halliwell criterion has a simple geometrical
interpretation. In terms of the real history states (\ref{realsplit})
the DHC can be written (ignoring null histories\footnote{A null
history is one with probability equal to $0$.})
\begin{eqnarray} \label{dhconh}
  \frac{|{\bf v}_{\alpha}^{T}{\bf v}_\beta|}{||{\bf
      v}_{\alpha}||\,||{\bf v}_\beta||} = |\cos(\theta_{\alpha\beta})|
      & \leq & \epsilon \quad \forall\, \alpha\neq\beta,
\end{eqnarray}
where $\theta_{\alpha\beta}$ is the angle between the real history
vectors ${\bf v}_{\alpha}$ and ${\bf v}_\beta$. The DHC requires that
the angle between every pair of histories must be at least
$\cos^{-1}\epsilon$ degrees.

In a $d$ dimensional Hilbert space there can only be $2d$ exactly
consistent, non-null histories. Thus, if a set contains more than $2d$
non-null histories, it cannot be continuously related to an exactly
consistent set unless some of the histories become null.  Establishing
the maximum number of histories satisfying (\ref{dhconh}) in finite
dimensional spaces is a particular case from a family of problems,
which has received considerable study.

\subsection{Generalised Kissing Problem}

The Generalised Kissing Problem is the problem of determining how many
$(k-1)$-spheres of radius $r$ can be placed on the surface of a sphere
with radius $R$ in ${\bf R}^k$.  This problem is equivalent to
calculating the maximum number of points that can be found on the
sphere all at least $\cos^{-1}\epsilon$ degrees apart, where $\epsilon
= 1 - 2r^2(R+r)^{-2}$.

To express these ideas mathematically, I define $M({\bf L}, ({\bf u},
{\bf v}) \leq s)$ to be the size of the largest subset of ${\bf L}$,
such that $({\bf u}, {\bf v}) \leq s$ for all different elements in
the subset, where ${\bf L}$ is a metric space. The Generalised Kissing
Problem is calculating
\begin{equation}
  M({\bf S}^{k-1},\, {\bf u}^T{\bf v} \leq \epsilon),
\end{equation}
where ${\bf S}^{k-1}$ is the set of points on the unit sphere in ${\bf
R}^{k}$.  The greatest number of histories satisfying the DHC is
\begin{equation}
  M({\bf CS}^{d-1},\, |\mbox{Re}({\bf u}^\dagger{\bf v})| \leq
  \epsilon) \,=\, M({\bf S}^{2d-1},\, |{\bf u}^T{\bf v}| \leq
  \epsilon)
\end{equation}
and for the medium DHC is
\begin{equation}
  M({\bf CS}^{d-1},\, |{\bf u}^\dagger{\bf v}| \leq \epsilon ),
\end{equation}
where ${\bf CS}^{d-1}$ is the set of points on the unit sphere in
${\bf C}^{d}$.

There is a large literature devoted to sphere-packing. Although few
exact results emerge from this work, numerous methods exist for
generating bounds. The tightest upper bounds derive from an
optimisation problem. In appendix (\ref{upperboundappendix}) I prove
that the well known bound
\begin{equation}\label{upperbound}
  M({\bf S}^{2d-1},\, |{\bf u}^T{\bf v}| \leq \epsilon) \leq
  \left\lfloor\frac{2d(1-\epsilon^2)}{1-2d\epsilon^2}\right\rfloor
\end{equation}
is the solution to the optimisation problem when $\epsilon^2 \leq
1/(2d+2)$.

The most important feature of this bound is that for $ 0 \leq \epsilon
\leq 1/(2d)$ it is exact, since for $\epsilon < 1/(2d) $ it gives $2d$
as an upper bound and for $\epsilon =1/(2d)$ it gives $2d +1$, and
there are packings that achieve these bounds\footnote{A packing with
$\epsilon =1/(2d)$ is generated by the rays passing through the $2d+1$
vertices of the regular $(2d)$-simplex.}. This is also the range of
most interest in Consistent Histories since an exactly consistent set
cannot contain more than $2d$ non-null histories. This result shows
that if $\epsilon < 1/(2d)$ then there cannot be more than $2d$
histories in a set satisfying the DHC.  Deciding when a set of vectors
could be a set of histories is a difficult problem, so this result
does not prove that this bound is optimal, although it is suggestive.

This bound (\ref{upperbound}) can now be used to prove several upper
bounds on probability sum rules.

\begin{eqnarray}
  \left|\sum_{\alpha \neq \beta \in \overline \alpha}
    D_{\alpha\beta}\right| & \leq & \sum_{\alpha \neq \beta \in
    \overline\alpha} |D_{\alpha\beta}|, \\ & \leq & \epsilon
    \sum_{\alpha \neq \beta \in \overline \alpha } \left(
    D_{\alpha\alpha} D_{\beta\beta} \right)^{1/2},\\ & \leq & \epsilon
    (n-1) \sum_{\alpha \in \overline \alpha }
    D_{\alpha\alpha}.\label{abc}
\end{eqnarray}
But the number of histories $n$ is bounded by
$2d(1-\epsilon^2)/(1-2d\epsilon^2)$, so
\begin{eqnarray}
  \left|\sum_{\alpha \neq \beta \in \overline \alpha}
    D_{\alpha\beta}\right|&\leq& \epsilon \,
    \frac{2d-1}{1-2d\epsilon^2}\sum_{\alpha \in \overline \alpha }
    D_{\alpha\alpha}.\label{DHsumlawbound}
\end{eqnarray}
Choose
\begin{equation}\label{an1}
  \epsilon(\delta) =
  \frac{-(2d-1)+\sqrt{(2d-1)^2+8d\delta^2}}{4d\delta}
\end{equation}
and then (\ref{DHsumlawbound}) implies
\begin{eqnarray}
  \left|\sum_{\alpha \neq \beta \in \overline \alpha} D_{\alpha\beta}
    \right| & \leq & \delta \sum_{\alpha \in \overline \alpha}
    D_{\alpha\alpha}.
\end{eqnarray}
This is the exact version of Dowker and Halliwell's result
(\ref{DHsumlaw}). For homogenous histories $\sum_{\alpha}
D_{\alpha\alpha} =1$ and then (\ref{DHsumlawbound}) and (\ref{an1})
imply
\begin{equation}\label{mainmpvbound}
  \mbox{MPV} < \delta.
\end{equation}
These results can easily be extended to general class-operators since
the same methods lead to a bound on $\sum_{\alpha} D_{\alpha\alpha}$
in terms of $\epsilon$.
\begin{eqnarray}
  \sum_{\alpha,\beta} D_{\alpha\beta} &=& 1\\ \Rightarrow
  \sum_{\alpha} D_{\alpha\alpha} &=& 1 -\sum_{\alpha \neq \beta}
  D_{\alpha\beta} \\ \Rightarrow \sum_{\alpha} D_{\alpha\alpha} & \leq
  & 1 + \sum_{\alpha \neq \beta}|D_{\alpha\beta}|\\ &\leq&1 + \epsilon
  (n-1) \sum_{\alpha} D_{\alpha\alpha} \\ \Rightarrow \sum_{\alpha}
  D_{\alpha\alpha} &\leq&\frac{1}{1- (n-1)\epsilon}.
\end{eqnarray}
There are sets of histories for which this bound is obtained. In
particular if $\epsilon = 1/(n-1)$ there are finite sets for which
$\sum_{\alpha} D_{\alpha\alpha}$ is arbitrarily large.  Inserting this
result into (\ref{DHsumlawbound}) results in
\begin{eqnarray}
  \left|\sum_{\alpha \neq \beta \in \overline \alpha}
    D_{\alpha\beta}\right| & \leq & \frac{\epsilon (n-1)}{1-
    (n-1)\epsilon}\\ & \leq & \epsilon
    \frac{2d-1}{1+\epsilon-2d\epsilon(1+\epsilon)}.\label{gb}
\end{eqnarray}
Choose
\begin{equation}\label{an2}
  \epsilon(\delta) = \frac{-(2d-1)(1+\delta) +\sqrt{(2d-1)^2
      (1+\delta)^2+ 8d\delta^2}} {4d\delta},
\end{equation}
and then (\ref{gb}) becomes
\begin{equation} 
  \left|\sum_{\alpha \neq \beta \in \overline \alpha}
    D_{\alpha\beta}\right| \leq \delta,
\end{equation}
so
\begin{equation}
  MPV \leq \delta.
\end{equation}
For physical situation the probability violation must be small so
$\delta << 1$ and these results can be simplified. From
(\ref{upperbound}) if $\epsilon < 1/(2d)$ $n\leq2d$ so (\ref{an1}) and
(\ref{an2}) can be simplified to
\begin{equation}\label{epschoice}
  \epsilon(\delta) = \frac{\delta}{2d}, \delta<1 \Rightarrow MPV \leq
  \delta + O(\delta^2),
\end{equation}
for all types of histories.  This is the main result of this paper. If
the medium DHC holds {\em or\/} the class-operators are homogenous
then (\ref{epschoice}) can be weakened to $\epsilon(\delta) =
\delta/d$ and still imply (\ref{mainmpvbound}). If the medium DHC
holds {\em and\/} the class-operators are homogenous then
(\ref{epschoice}) can be further weakened to $\epsilon(\delta) =
2\delta/d$ and still imply (\ref{mainmpvbound}).

In appendix (\ref{dheg}) I give a simple example of a family of sets
of histories, of any size, satisfying the medium DHC with $\mbox{MPV}
= d\epsilon/4$. If $\epsilon$ is chosen according to (\ref{epschoice})
then the $\mbox{MPV} = \delta/8$.  This example illustrates that
equation (\ref{epschoice}) is close to the optimal bound.
Since the example satisfies the {\em medium\/} DHC and the
class-operators are homogenous $\epsilon$ can be chosen to be
$2\delta/d$ and the MPV is then $\delta/2$, so for this example the
bound is achieved within a factor of two.

The choice $\epsilon = \delta/(2d)$ in relation to the DHC is
particularly convenient in computer models. Often one constructs a set
of histories by individually making projections, and one desires a
simple criterion which will bound the MPV. The DHC solves this
problem.

The only known lower bounds for the generalised kissing problem derive
{}from an argument of Shannon's \cite{Shannon} developed by
Wyner\cite{Wyner}. Shannon proved
 
\begin{eqnarray} \label{slbound}
  M({\bf S}^{2d-1},\, |{\bf u}^T{\bf v}| \leq \epsilon) & \geq &
  (1-\epsilon^2)^{1/2-d}.
\end{eqnarray}
I explain the proof and extend it for the medium DHC in appendix
(\ref{shannonap}).

This simple bound (\ref{slbound}) has an important consequence: the
number of histories satisfying the DHC can increase exponentially with
$d$ if $\epsilon$ is constant. So for constant $\epsilon > 0$ by
choosing a large enough Hilbert space the MPV can be arbitrarily
large, therefore $\epsilon$ must be chosen according to the dimension
of the Hilbert space.

When the Hilbert space is infinite-dimensional and separable, and
$\epsilon > 0$, (\ref{slbound}) suggests that there can be an
uncountable number of histories satisfying the DHC. If so, the DHC can
only guarantee proximity to an exactly consistent set for finite
Hilbert spaces. Though if the system is set up in a Hilbert space of
dimension $d$ and the limit $d\to\infty$, $\epsilon =
O(\sqrt{\frac{\log d}{d}})$ is taken (assuming it exists) then the
bound remains countable and it may be useful even for infinite spaces.
 
If there are $n$ histories satisfying the DHC with $\epsilon =
\delta/(n-1)$, then, from (\ref{abc}), $\mbox{MPV} \leq \delta + O(
\delta^2)$. This result is trivial, but the DHC also ensures that the
histories will span a subspace of dimension at least $n/2$. Therefore,
there will be exactly consistent sets with the same number of non-null
histories that span the same subspace.

If
\begin{eqnarray*}
  \epsilon \leq \left[1-(2d)^{2/(1-2d)}\right]^{1/2} & = &
    \left[\frac{2\ln 2d}{2d-1}\right]^{1/2} \\&&\mbox{}+ O\left\{
    \left[\frac{\ln d}{d}\right]^{3/2}\right\}
\end{eqnarray*} 
then the lower bound is less than the trivial lower bound $M \geq d$.
Since the upper bounds (\ref{upperbound}) holds only for $\epsilon
\leq O(1/\sqrt(d))$ the two sets of bounds are not mutually useful.
The Shannon bound is too poor for small $\epsilon$ because it ignores
the overlap between spherical caps. A more rigorous bound would add
points one by one on the edge of existing caps, and allow for the
overlap between them. Unfortunately, there are no useful results in
this direction.

\subsection{Discontinuity of the Dowker-Halliwell Criterion}

A consistent set of histories can be extended to another consistent
set by repeating projector sets at adjacent times. Moreover the two
sets of histories are physically equivalent descriptions. Both the
preceding statements are true because the class-operators for the two
sets are identical, since $P_iP_j = \delta_{ij}P_i$. If a slightly
perturbed set of projectors is used the result is more complicated.

For simplicity I shall consider a simple example consisting of a set
of $n$ history states $\{{\bf u}_\alpha\}$, and a projector $P$ and
its complement $\overline{P}$.  This is sufficiently general to deal
with all cases.  Suppose that the set of histories is consistent when
extended by $\{P,\overline{P}\}$, so
\begin{eqnarray}\label{phs}
  \mbox{Re} ( {\bf u}_\alpha^\dagger P {\bf u}_\beta) & = & 0, \\
  \mbox{Re} ( {\bf u}_\alpha^\dagger \overline{P} {\bf u}_\beta) & = &
  0.
\end{eqnarray}
Then the set extended again by $\{P,\overline{P}\}$ will trivially be
consistent, $2n$ of the histories will be unaltered, the other $2n$
will be null.  Now suppose the projectors are slightly perturbed
without altering their ranks. The perturbed projectors will be related
to the old ones by a unitary transformation $P' = U(\epsilon)^\dagger
P U(\epsilon)$, and $U(\epsilon)$ can be written $U(\epsilon) =
\exp{(i\epsilon A)}$, where $A$ is Hermitian. The set of histories
$\{P'P{\bf u}_\alpha, P'\overline{P}{\bf u}_\alpha, \overline{P'}P{\bf
u}_\alpha, \overline{P'}\overline{P}{\bf u}_\alpha\}$ will no longer
necessarily be consistent, but it is approximately consistent to
$O(\epsilon)$ since it has only been perturbed to this order from an
exactly consistent set. To second order in $\epsilon$
\begin{eqnarray}\label{pops}
  PP'P & = & P - \epsilon^2PA\overline{P}AP, \\ \overline{P}P'P & = &
  -i\epsilon\overline{P}AP +\epsilon^2\overline{P}A(P - 1/2)AP, \\
  \overline{P}P'\overline{P} & = &
  \epsilon^2\overline{P}APA\overline{P}.
\end{eqnarray}
Using (\ref{phs}) and (\ref{pops}) the real part of the new
decoherence matrix to leading order in $\epsilon$ contains blocks like
\widetext
\begin{equation}
  \mbox{Re}\left(
    \begin{array}{cccc}
           {\bf u}_\alpha^\dagger P{\bf u}_\alpha & i \epsilon{\bf
u}_\alpha^\dagger PA\overline{P} {\bf u}_\alpha & -\epsilon^2{\bf
u}_\alpha^\dagger PA\overline{P}AP {\bf u}_\beta & i \epsilon{\bf
u}_\alpha^\dagger PA\overline{P} {\bf u}_\beta \\ -i \epsilon{\bf
u}_\alpha^\dagger \overline{P}AP {\bf u}_\alpha & \epsilon^2{\bf
u}_\alpha^\dagger \overline{P}APA\overline{P} {\bf u}_\alpha &
-i\epsilon{\bf u}_\alpha^\dagger \overline{P}AP {\bf u}_\beta &
\epsilon^2{\bf u}_\alpha^\dagger \overline{P}APA\overline{P} {\bf
u}_\beta \\ -\epsilon^2{\bf u}_\beta^\dagger PA\overline{P}AP {\bf
u}_\alpha & i\epsilon{\bf u}_\beta^\dagger PA\overline{P} {\bf
u}_\alpha & {\bf u}_\beta^\dagger P {\bf u}_\beta & i\epsilon{\bf
u}_\beta^\dagger PA\overline{P} {\bf u}_\beta \\ -i\epsilon{\bf
u}_\beta^\dagger \overline{P}AP {\bf u}_\alpha & \epsilon^2{\bf
u}_\beta^\dagger \overline{P}APA\overline{P} {\bf u}_\alpha &
-i\epsilon{\bf u}_\beta^\dagger \overline{P}AP {\bf u}_\beta &
\epsilon^2{\bf u}_\beta^\dagger \overline{P}APA\overline{P} {\bf
u}_\beta
    \end{array}
  \right).
\end{equation}
\narrowtext

All off-diagonal terms are at least as small as $O(\epsilon)$, but the
 DHC terms will be

\begin{eqnarray}\label{dhcterms}
  \frac{|\mbox{Im}({\bf u}_\alpha^\dagger PA\overline{P} {\bf
      u}_\beta)|} {||P {\bf u}_\alpha||\,||PA\overline{P} {\bf
      u}_\beta||} & \quad \mbox{and} \quad & \frac{|\mbox{Re}({\bf
      u}_\alpha^\dagger\overline{P}A PA\overline{P} {\bf u}_\beta)|}
      {||PA\overline{P} {\bf u}_\alpha||\,||PA\overline{P} {\bf
      u}_\beta||},
\end{eqnarray}
which are $O(1)$, and not necessarily small. These terms
(\ref{dhcterms}) are {\em discontinuous\/} as $\epsilon$ varies,
because for $\epsilon = 0$ they are ill defined since they correspond
to $0/0$, and for $\epsilon > 0$ they can take any value between $0$
and $1$. Since these terms are the overlap of two unit vectors in a
$\mbox{Rank}(P)$ dimensional space, the primary determining factor is
the rank of $P$.

To proceed further one must make assumptions about $A$. If one wants
to consider a random perturbation a natural requirement is that there
is no preferred direction, hence $A$ is drawn from a distribution
invariant under the unitary group. Random Hermitian matrices of this
form have been much studied \cite{Mehta}. Approximate expectations can
then be calculated for the terms (\ref{dhcterms}) and are
$[\mbox{Rank}(P)]^{-1/2}$ which will always be much larger than
$1/(2n)$. Since {\em every\/} off-diagonal element must satisfy the
DHC, only in exceptional cases will a slightly perturbed set of
projectors lead to a set that satisfy the DHC.  This effect occurs
because the histories that are null in the limit of exact consistency
(as $\overline{P}P' \to 0$) now have a finite, though very small,
probability.

This is a useful feature as these almost-null histories are
uninteresting. Only if there is some special relation between the
states, the projectors and the perturbation will the new set be
consistent, in which case it is adding information.

The exception to the above occurs if a {\em single\/}, binary,
branch-dependent projection set is used, and {\em all\/} the histories
lie in the null space of one of the projectors. Use the projection set
$\{P',\overline{P'}\}$ on the history ${\bf u}_1$ when
$\overline{P}{\bf u}_\alpha = 0 $ for all $\alpha$. Then the real part
of the new decoherence matrix to leading order in $\epsilon$ contains
blocks like
\begin{equation}
  \mbox{Re}\left(
  \begin{array}{cccc}
    {\bf u}_1^\dagger P{\bf u}_1 & 0 & -\epsilon^2{\bf u}_1^\dagger
    A\overline{P}A {\bf u}_\alpha \\ 0 & \epsilon^2{\bf u}_1^\dagger
    A\overline{P}A{\bf u}_1 & \epsilon^2{\bf u}_1^\dagger
    A\overline{P}A{\bf u}_\alpha \\ -\epsilon^2{\bf u}_\alpha^\dagger
    A\overline{P}A{\bf u}_1 & \epsilon^2{\bf u}_\alpha^\dagger
    A\overline{P}A{\bf u}_\alpha & {\bf u}_\alpha^\dagger {\bf
    u}_\alpha
  \end{array}
\right).
\end{equation}
All the off-diagonal terms are $O(\epsilon^2)$ and all the diagonal
terms are $O(1)$, except for one which is $O(\epsilon^2)$, so this set
satisfies the DHC to $O(\epsilon)$. This is a very special case and is
unlikely to occur in real examples.

\subsection{Other Properties of the Dowker-Halliwell Criterion}
In standard Quantum Mechanics the probability of observing a system in
state $|\phi\rangle$ when it is in state $|\psi\rangle$ is
$|\langle\phi|\psi\rangle|^2 / (\langle\phi|\phi\rangle \,
\langle\psi|\psi\rangle)$. If we take this as a measure of
distinguishability then the set of history states, $\{{\bf
u}_\alpha\}$, are distinguishable to order $\epsilon^2$ only if
\begin{equation}
  \frac{|{{\bf u}_\alpha}^\dagger {{\bf u}_\beta}|^2} {||{\bf
    u}_\alpha||^2||{\bf u}_\beta||^2} \, \leq \, \epsilon^2 \quad
    \forall \alpha \neq \beta.
\end{equation}
But this is equivalent to the medium DHC (\ref{DHC}) since
\begin{equation}
  \frac{|{{\bf u}_\alpha}^\dagger {\bf u}_\beta|} {||{\bf
  u}_\alpha||\,||{\bf u}_\beta||} = \frac{|D_{\alpha\beta}|}
  {(D_{\alpha\alpha} D_{\beta\beta})^{1/2}}.
\end{equation}
Histories which only satisfy the weak consistency criterion
(\ref{weakcon}) need not be distinguishable since a pair of histories
may only differ by a factor of $i$ and would be regarded as equivalent
in conventional quantum mechanics. This is one of the few differences
between the medium (\ref{MDHC}) and the standard (\ref{DHC}) DHC.

Outside of quantum cosmology one usually discusses conditional
probabilities: one regards the past history of the universe as
definite and estimates probabilities for the future from it. One does
this in consistent histories by forming the {\em current density
matrix \/} $\rho_c$. Let $\{C_\alpha\}$ be a complete set of
class-operators, each of which can be divided into the the past and
the future, $C_\alpha = C^f_{\alpha_f}C^p_{\alpha_p}$. Then the
probability of history $\alpha_f$ occuring given $\alpha_p$ has
occurred is
\begin{eqnarray} 
  P(\alpha_f | \alpha_p) & = & \frac{P(\alpha_f \, \& \,
    \alpha_p)}{P(\alpha_p)}, \\ & = & \frac{\mbox{Tr} (C^f_{\alpha_f}
    C^p_{\alpha_p} \rho C_{\alpha_p}^{p\dagger}
    C_{\alpha_f}^{f\dagger})}{\mbox{Tr} (C^p_{\alpha_p} \rho
    C_{\alpha_p}^{p\dagger})}, \\ & = & \mbox{Tr} (C^f_{\alpha_f}
    \rho_c C_{\alpha_f}^{f\dagger})\label{cdm},
\end{eqnarray}
where $\rho_c = C^p_{\alpha_p} \rho C_{\alpha_p}^{p\dagger} /
\mbox{Tr} (C^p_{\alpha_p} \rho C_{\alpha_p}^{p\dagger})$. Equation
(\ref{cdm}) shows that all future probabilities can be expressed in
terms of $\rho_c$. The DHC in terms of $\rho_c$ is
\begin{equation}\label{dhcb}
  \frac{\mbox{Tr} (C^f_{\gamma_f} \rho_c C_{\beta_f}^{f\dagger})}
  {[\mbox{Tr} (C^f_{\gamma_f} \rho_c C_{\gamma_f}^{f\dagger})
  \mbox{Tr} (C^f_{\beta_f} \rho_c C_{\beta_f}^{f\dagger})]^{1/2}} \leq
  \epsilon, \quad \forall \gamma_f \neq \beta_f.
\end{equation}
This is the same as the DHC applied to the complete histories, given
the past,
\begin{equation}\label{dhca}
  \frac{\mbox{Tr} (C^f_{\gamma_f} C^p_{\alpha_p} \rho
    C_{\alpha_p}^{p\dagger} C_{\beta_f}^{f\dagger})} {[\mbox{Tr}
    (C^f_{\gamma_f} C^p_{\alpha_p} \rho
    C_{\alpha_p}^{p\dagger}C_{\gamma_f}^{f\dagger}) \mbox{Tr}
    (C^f_{\beta_f} C^p_{\alpha_p} \rho
    C_{\alpha_p}^{p\dagger}C_{\beta_f}^{f\dagger})]^{1/2}} \leq
    \epsilon,
\end{equation}
for all $\gamma_f \neq \beta_f$.  This is a property not possessed by
the usual criterion (\ref{badcriterion}) or any other based on
absolute probabilities, such as one that only bounds the MPV. This is
an important property since any non-trivial\footnote{i.e. a branch
whose probability is non-zero.} branch of a consistent set of
histories (when regarded as a set of histories in its own right) must
also be consistent and one would like a criterion for approximate
consistency that reflects this.

Experiments in quantum mechanics are usually carried out many times,
and the relative frequencies of the outcomes checked with their
probabilities predicted by quantum mechanics. Consider the situation
where an experiment is carried out at $m$ times $\{t_i\}$ with
probabilities $\{p_i\}$. Let $P^i$ be the projector corresponding to
the experiment being performed at $t_i$ and let
$\{C_\alpha^i\}=\{U(-t_i)C_\alpha U(t_i)\}$ be the set of $n$
class-operators corresponding to the different outcomes of the
experiment when it is started at time $t_i$. For simplicity assume
that the probability of an experiment being performed and its results
are independent of other events. This implies $[P^i,P^j] = 0$,
$[P^i,C^j_\alpha] = 0$ and $[C^i_\alpha,C^j_\alpha] = 0$ so $p_i =
\langle\phi|P^i|\phi\rangle$. There are $(1+n)^m$ class-operators and
they are of the form
\begin{equation}\label{egco}
  \overline{P^{i_k}}\ldots\overline{P^{i_1}} \, P^{j_{m-k}} \ldots
  P^{j_1} \, C_{\alpha_{m-k}}^{j_{m-k}} \ldots C_{\alpha_1}^{j_1},
\end{equation}
corresponding to the experiment being performed at times $t_{j_1}
\ldots t_{j_m-k}$ and not at times $t_{i_1} \ldots t_{i_k}$ with
results $\alpha_1 \ldots \alpha_{m-k}$. Because of the commutation
relations the only non-zero off-diagonal-elements of the decoherence
matrix contain factors like
\begin{equation}\label{nzod}
  p_i\,\mbox{Re}(\langle\psi|C_\beta^\dagger C_\alpha|\psi\rangle),
\end{equation}
where $|\psi\rangle$ is the initial state in which the experiment is
prepared identically each time. When the environment and time between
experiments are large the $P_i$ will commute and this justifies the
usual arguments where the consistency of the experiment alone is
considered rather than the consistency of the entire run of
experiments.

This is a particular case of the result that an inconsistent set
cannot extend a non-trivial branch of a set of histories without
destroying its consistency. A sensible criterion for approximate
consistency should also have this property. By choosing the $p_i$
small enough the off-diagonal elements (\ref{nzod}) can be made
arbitrarily small, thus any criterion for approximate consistency
which uses absolute probabilities will regard the set as consistent,
however inconsistent the experiment itself may be.

An important feature of the DHC is that it has no such disadvantage,
as the $p_i$'s will cancel and the approximate consistency conditions
will be
\begin{equation}
  \frac{|\mbox{Re}(\langle\psi|C_\beta^\dagger
    C_\alpha|\psi\rangle)|}{||C_\alpha|\psi\rangle||\,
    ||C_\beta|\psi\rangle||} \leq \epsilon.
\end{equation}

\section{Conclusions}

A set of histories is approximately consistent to order $\delta$, only
if its MPV is less than $\delta$. The often-used criterion
\begin{equation} \label{ccagain}
  \mbox{Re}(D_{\alpha\beta}) \leq \epsilon(\delta), \quad \forall
  \alpha \neq \beta
\end{equation}
is not sufficient for any $\epsilon(\delta) > 0$, since there are sets
of histories satisfying (\ref{ccagain}) with arbitrarily large MPV.
The criterion (\ref{ccagain}) can only be used if $\epsilon(\delta) =
O(1/n^2)$, where $n$ is the number of histories. The Dowker-Halliwell
criterion has no such disadvantage.

If
\begin{equation}\label{tc}
  \mbox{Re}(D_{\alpha\beta}) \leq \frac{\delta}{2d}\,
  (D_{\alpha\alpha}D_{\beta\beta})^{1/2} \quad \mbox{$\forall \alpha
  \neq \beta$, $\delta < 1$,}
\end{equation}

then the MPV is less than $\delta + O(\delta^2) $. This is the paper's
main result.  If the medium DHC holds,
\begin{equation}\label{mtc}
  |D_{\alpha\beta}| \leq \frac{\delta}{d}\,
  (D_{\alpha\alpha}D_{\beta\beta})^{1/2} \quad \mbox{$\forall \alpha
  \neq \beta$, $\delta < 1$,}
\end{equation}
then the MPV is also bounded by $\delta$. For histories satisfying
either criterion, if only homogenous class-operators are used then the
upper bound on the MPV is strengthened to $\delta/2$.  The bounds are
also optimal in the sense that they are can be achieved (to within a
small factor) in any finite dimensional Hilbert space.  Any improved
bound must use the global structure of the decoherence matrix.

The DHC is particularly suitable for computer models in which a set of
histories is built up by repeated projections. If each history
satisfies (\ref{tc}) as it is added, then the whole set will be
consistent to order $\delta$ and there will be no more than $2d$
histories.

The DHC also leads to a simple, geometrical picture of consistency:
the path-projected states can be regarded as pairs of points on the
surface of a hyper-sphere, all separated by at least
$\cos^{-1}\epsilon$ degrees. This approach can be used to prove that
$\epsilon$ in the DHC must be chosen according to the dimension of the
Hilbert space. Ideally one would like a criterion for approximate
consistency that implied the existence of an exactly consistent set
corresponding to physical events that only differed to order
$\epsilon$. The DHC seems well adapted to defining proximity to an
exactly consistent set and may be useful in constructing a proof that
such a set exists.

This bound (\ref{slbound}) shows that the number of histories
satisfying the DHC can increase exponentially with $d$ if $\epsilon$
is constant. So for constant $\epsilon > 0$ by choosing a large enough
Hilbert space the MPV can be arbitrarily large, therefore $\epsilon$
must be chosen according to the dimension of the Hilbert space.

If a set is not exactly consistent then it cannot be a subset of an
exactly consistent set (unless the branch is trivial.) The same is
true for approximate consistency when it is defined by the
DHC. However, this is not true however for any criterion which depends
solely on MPV. It is a particularly useful property when discussing
conditional probabilities.

\section{acknowledgements}
This work is supported by a studentship from the United Kingdom
Engineering and Physical Sciences Research Council. I would like to
thank Adrian Kent for his generous support and constructive
criticisms, and Fay Dowker and Jonathan Halliwell for helpful
comments.

\appendix

\section{Sphere-packing Bounds}\label{upperboundappendix}
\subsection{Upper Bounds Using Zonal Spherical Harmonic Polynomials} 
Various authors \cite{Kabatyanski:Levenshtein,Delsarte} have
constructed upper bounds for $M$ by using the properties of zonal
spherical harmonic polynomials, which for many spaces are the Jacobi
polynomials $P_n^{(\alpha,\beta)}(x)$. The bounds
\begin{equation}
  \label{Delsarte:bound}
  M({\bf S}^{d-1},\, |{\bf u}^T{\bf v}| \leq \epsilon) = N((d-3)/2,
  -1/2, 2\epsilon^2-1),
\end{equation}
for $d \geq 3$, and
\begin{equation} 
  M({\bf CS}^{d-1},\, |{\bf u}^\dagger{\bf v}| \leq \epsilon) = N(d-2,
  0, 2\epsilon^2-1),
\end{equation}
for $d \geq 2$, have been proved by Kabatyanski et
al. \cite{Kabatyanski:Levenshtein} and (\ref{Delsarte:bound}) also by
Delsarte et al.  \cite{Delsarte}.  Here $N(\alpha, \beta, s)$ is
defined as the solution to the following optimisation problem.

Consider $s$ as a given number $-1 \leq s < 1$. Let ${\cal R}
(\alpha,\beta,s)$ be the set of polynomials of degree at most $k$ with
the following properties:
\begin{eqnarray*}
  f(t) & = & \sum_{i=0}^k f_i P_i^{(\alpha,\beta)}(t), \\ f_i & \geq &
  0, \quad i = 0,1,\ldots,k, \quad \mbox{and} \quad f_0 > 0,\\ f(t) &
  \leq & 0 \quad \mbox{for} \quad -1 \leq t \leq s.
\end{eqnarray*}
Then
\begin{displaymath}
  N(\alpha, \beta, s) = \inf_{f(t) \in {\cal R}(\alpha, \beta, s)}
  f(1)/f_0.
\end{displaymath}
This can be converted to a linear program by defining
\begin{displaymath}
  \tilde{P}_i^{(\alpha,\beta)}(t) = P_i^{(\alpha,\beta)}(t) /
  P_i^{(\alpha,\beta)}(1).
\end{displaymath}
Then $ N(\alpha, \beta, s) = 1 + \sum_{i=1}^k f_i, $ where $
\sum_{i=1}^k f_i $ is minimised subject to $ f_i \geq 0 $ and $
\sum_{i=1}^k f_i \tilde{P}_i^{(\alpha,\beta)}(t) \leq -1, $
for $-1 \leq t \leq s$.
This formulation is discussed in Conway and Sloane \cite{Conway}, but
no exact solutions are known. However, any $f(t)$ satisfying the
constraints does provide a bound, though it may not be optimal. I show
in appendix \ref{alpha:half} that
\begin{displaymath}
  \tilde{P}^{(\alpha,-1/2)}_n(x) > \tilde{P}^{(\alpha,-1/2)}_1(x),
\end{displaymath}
if $-1 < x < -(2\alpha+3)(2\alpha+5)^{-1}$, $n > 1$ and $\alpha \geq
1$. So if $s$ is less than $-(2\alpha+3)/(2\alpha+5)$ then
$\tilde{P}^{(\alpha,-1/2)}_1(t)$ is more negative than any other of
the $\tilde{P}^{(\alpha,-1/2)}_i(t)$, and since
$\tilde{P}^{(\alpha,-1/2)}_1(t)$ is increasing the solution is
\begin{displaymath} 
  f_i = 0, \quad i = 2,3,\ldots,\quad f_1 =
  -1/\tilde{P}^{(\alpha,-1/2)}_1(s) \quad \forall k.
\end{displaymath}
So the optimal bound using zonal spherical harmonics is
\begin{eqnarray}\nonumber
  M({\bf CS}^{d-1}, \mbox{Re}({\bf u^\dagger v}) \leq \epsilon) & = &
  M({\bf S}^{2d-1},\, |{\bf u}^T{\bf v}| \leq \epsilon) \\\nonumber &
  \leq & N(d-3/2, -1/2, 2\epsilon^2-1), \\\nonumber & = & 1 -
  1/\tilde{P}^{(d-3/2,-1/2)}_1(2\epsilon^2-1) \\ \label{R:upper:bound}
  & = & \frac{2d(1-\epsilon^2)}{1-2d\epsilon^2},
\end{eqnarray}
if $\epsilon^2 \leq 1/(2d+2)$ and $d \geq 3$.  I prove a similar
inequality in appendix \ref{alpha:zero} for $\alpha = 0$. So for the
medium DHC
\begin{eqnarray}\nonumber
  M({\bf CS}^{d-1}, |{\bf u^\dagger v}| \leq \epsilon) & \leq & N(d-2,
  0, 2\epsilon^2-1),\\ & = &
\label{M:upper:bound}
\frac{d(1-\epsilon^2)}{1-d\epsilon^2},
\end{eqnarray}
if $\epsilon^2 \leq 1/(d+1)$ and $d \geq 2$.

\subsection{Shannon's Lower Bound}\label{shannonap}
In a pioneering paper\cite{Shannon} Shannon proved

{\bf Theorem 1}
\begin{equation}
  M({\bf S}^{d-1},|{\bf u}^T{\bf v}| \leq \cos\theta) \geq
  \sin^{1-d}\theta.
\end{equation}

Let
\begin{equation} 
  S_d(r) = d r^{d-1}\pi^{d/2}/\Gamma[(d+2)/2]
\end{equation}
be the surface area of a sphere in Euclidean $d$-space of radius $r$,
and let $A_d(r,\theta)$ be the area of a $d$-dimensional spherical cap
cut from a sphere of radius $r$ with half angle $\theta$. It is not
hard to show that
\begin{equation} 
A_d (r,\theta) = \frac{(d-1)r^{d-1}\pi^{d-1/2}}{\Gamma[(d+2)/2]}
\int^{\theta}_{0}\sin^{d-2}\phi\, d \phi.
\end{equation}

Consider the largest possible set of rays through the origin
intersecting a sphere at points points ${\bf u} \in {\bf
S}^{d-1}$. About each point ${\bf u}$, consider the spherical cap of
all points on the sphere within $\theta$ degrees. Now, the set of all
such caps about each point ${\bf u}$ must cover the entire surface of
the sphere, otherwise we could add a new ray passing through the
uncovered areas. Since the area of each cap is $A_d(r,\theta)$, we
have
\begin{equation}
  2\, A_d(r,\theta) \,M({\bf S}^{d-1},|{\bf u}^T{\bf v}| \leq
  \cos\theta)\,\geq\, S_d(r) \mbox{.}
\end{equation}
But a spherical cap, $A_d(r,\theta)$, is contained within a hemisphere
of radius $r\sin\theta$, $A_d(r,\theta) \leq 1/2\,S_d(r\sin\theta)$
\footnote{This is easy to prove by changing variables in the integral
to $\sin\phi = \sin\theta\,\sin\psi$} , so
\begin{equation}
  M({\bf S}^{d-1},|{\bf u}^T{\bf v}| \leq \cos\theta) \geq
  S_d(r)/S_d(r\sin\theta) = \sin^{1-d}\theta
  \,\,{\rule[0ex]{1.5ex}{1.5ex}}
\end{equation}
or
\begin{equation}
  M({\bf CS}^{d-1},\mbox{Re}({\bf u^\dagger v)} \leq \cos\theta) \geq
  \sin^{1-2d}\theta.
\end{equation}

The straightforward extension of the proof to the complex case does
not appear to exist in the literature. It is slightly simpler as it is
easy to calculate the integral $A_d(r,\theta)$ exactly.

{\bf Theorem 2} {\em
  \begin{equation}
    M({\bf CS}^{d-1},|{\bf u^\dagger v}| \leq \cos\theta) \geq
    \sin^{2-2d}\theta
  \end{equation}
  }

The area of a unit sphere in ${\bf CS}^{d-1}$ is $S_{2d}(1)$. Let
$A_d(1,\theta)$ now be the area of a cap defined by
\begin{equation}
  \{{\bf u} \in {\bf CS}^{d-1} : |u_1|^2 \geq \cos\theta\}.
\end{equation}
We can choose coordinates for a vector ${\bf u}$ in ${\bf CS}^{d-1}$
by defining
\begin{eqnarray*}
  \mbox{Re}{(u_1)} & = & \cos\phi_1, \\ \mbox{Im}{(u_1)} & = &
  \sin\phi_1\,\cos\phi_2, \\ \vdots & \vdots & \vdots, \\
  \mbox{Re}{(u_d)} & = &
  \sin\phi_1\,\sin\phi_2\,\sin\phi_3\ldots\sin\phi_{2d-2}\cos\psi,\\
  \mbox{Im}{(u_d)} & = &
  \sin\phi_1\,\sin\phi_2\,\sin\phi_3\ldots\sin\phi_{2d-2}\sin\psi,
\end{eqnarray*}
where $\phi_n \in [0,\pi)$ and $\psi \in [0,2\pi)$.Then, by
integrating over $\phi_2,\phi_3,\ldots,\phi_{2d-2}$ and $\psi$, we get
\begin{eqnarray}\nonumber
  A_d(1,\theta) &=& S_{2d-2}(1)\,
  \!\!\!\!\!\!\!\!\!\!\!\!\!\!\!\!\!\!\!\!\!\!\!
  \mathop{\int\!\!\int}_{\cos^2\phi_1 +\sin^2\phi_1\cos^2\phi_2 \geq
  \cos\theta} \!\!\!\!\!\!\!\!\!\!\!\!\!\!\!\!\!\!\!\!\!\!\!
  \sin^{d-2}\phi_1\sin^{d-3}\phi_1\, d \phi_1\, d \phi_2\\ &=&
  \frac{\pi S_{2d-2}(1) \sin^{2d-2}\theta}{d-2}.
\end{eqnarray} 
Hence, using Shannon's argument again,
\begin{eqnarray}\nonumber
  M({\bf CS}^{d-1},|{\bf u^\dagger v}| \leq \cos\theta) &\geq& \frac
  {(d-2)\,S_{2d}(1)} {\pi S_{2d-2}(1)\sin^{2d-2}\theta}\\ &\geq&
  \sin^{2-2d}\theta \,\,{\rule[0ex]{1.5ex}{1.5ex}}
\end{eqnarray}

Expressed in terms of $\epsilon = \cos\theta$ the bounds are
\begin{eqnarray*}
  M({\bf CS}^{d-1}, \mbox{Re}({\bf u^\dagger v}) \leq \epsilon) & \geq
  & (1-\epsilon^2)^{1/2-d}, \\ \mbox{and} \qquad M({\bf CS}^{d-1},
  |{\bf u^\dagger v}| \leq \epsilon) & \geq & (1-\epsilon^2)^{1-d}.
\end{eqnarray*}

\section{Jacobi Polynomials}
I have used trivial properties of the Jacobi polynomials without
citation. All of these results can be found in chapter IV of Szeg\"o,
\cite{Szego} which provides an excellent introduction to, and
reference source for, the Jacobi polynomials.

\subsection{$S^{d-1}$, $\beta = -1/2$}\label{alpha:half}
In ${\bf S}^{d-1}$ the zonal spherical polynomials are
$P^{(\alpha,-1/2)}_n(x)$ with $\alpha = (d-3)/2$.

{\bf Theorem 3} {\em
  \begin{equation} 
    \label{inequalitybhalf}
    \tilde{P}^{(\alpha,-1/2)}_n(x) > \tilde{P}^{(\alpha,-1/2)}_1(x)
  \end{equation}
  for $-1<x<-(2\alpha+3)(2\alpha+5)^{-1}$, $n > 1$ and $\alpha \geq
  1$, where $\tilde{P}^{(\alpha,-1/2)}_n(x) =
  P^{(\alpha,-1/2)}_n(x)/P^{(\alpha,-1/2)}_n(1)$.  }

I begin by considering two special cases, $n = 2$ and $n = 3$. The
first four polynomials are:
\widetext
\begin{eqnarray*}
  \tilde{P}^{(\alpha,-1/2)}_0 (x) & = & 1 \\
  \tilde{P}^{(\alpha,-1/2)}_1 (x) & = & \frac{2\alpha + 1 + (2\alpha +
  3)x}{4(\alpha+1)}\\ \tilde{P}^{(\alpha,-1/2)}_2 (x) & = &
  \frac{4\alpha^2 - 13 + 2(2\alpha+1)(2\alpha+5)x
  +(2\alpha+5)(2\alpha+7)x^2}{ 16(\alpha+1)(\alpha+2)}\\
  \tilde{P}^{(\alpha,-1/2)}_3 (x) & = &
  \frac{(2\alpha+1)(4\alpha^2-8\alpha-57) +
    3(2\alpha+7)(4\alpha^2-21)x}{}\\&& 
  \frac{\mbox{} +3(2\alpha+1)(2\alpha+7)(2\alpha+9)x^2}{}\\&&
  \frac{\mbox{} +(2\alpha+7)(2\alpha+9)(2\alpha+11)x^3}{64(\alpha+1)
    (\alpha+2)(\alpha+3)}.
\end{eqnarray*}
So
\begin{eqnarray}\label{p2-p1-1/2} 
  \tilde{P}^{(\alpha,-1/2)}_2 (x) - \tilde{P}^{(\alpha,-1/2)}_1 (x)
  &=& \frac{-(2\alpha+7)(1-x)[2\alpha+3 + (2\alpha+5)x]}{
  16\,(\alpha+1)(\alpha+2)}
\end{eqnarray}
\begin{eqnarray}\nonumber
  \tilde{P}^{(\alpha,-1/2)}_3 (x) - \tilde{P}^{(\alpha,-1/2)}_1 (x)
  &=& \\ && \!\!\!\!\!\!\!\!\! \!\!\!
  \frac{-(2\alpha+9)(1-x)[(2\alpha+1)(6\alpha+17)+
    2\,(2\alpha+7)(4\alpha+7)x}{}\\&&
  \frac{\mbox{} + (2\alpha+7)(2\alpha+11)x^2]}{
    64\,(\alpha+1)(\alpha+2)(\alpha+3)}.
\label{p3-p1-1/2}
\end{eqnarray}
\narrowtext
Equation (\ref{p2-p1-1/2}) is positive for \mbox{$x
  <-(2\alpha+3)(2\alpha+5)^{-1} $} (hence the range chosen for
  (\ref{inequalitybhalf}).)  Equation (\ref{p3-p1-1/2}) is positive
  where the quadratic factor
\begin {eqnarray} \nonumber
  (2\alpha+1)(6\alpha+17) + 2(2\alpha+7)(4\alpha+7)x \\ \mbox{}
  +(2\alpha+7)(2\alpha+11)x^2 \label{quadratic}
\end{eqnarray}
is negative. Since (\ref{quadratic}) is positive for large $|x|$ if it
is negative at any two points it will be negative in between. At $x =
-1$ it is $-4\,(2\alpha+1)$, and at $x = -(2\alpha+3)(2\alpha+5)^{-1}$
it is $ -16\,(\alpha+2)(2\alpha+11)(5+2\alpha)^{-2}$, which is
negative for $\alpha > -2$.  So the inequality (\ref{inequalitybhalf})
holds for $n=2$ and $n=3$.

For $n > 3$ the inequality is easily proved, by bounding the solutions
of the Jacobi differential equation,
\begin{eqnarray}\nonumber
  (1-x^2) y''(x)+[\beta - \alpha - (\alpha+\beta+2)x] y'(x)\\ \mbox{}
  + n(n+\alpha+\beta+1) y(x) = 0,
\label{jacobi ode 1}
\end{eqnarray} 
where $y(x)=P^{(\alpha,\beta)}_n(x)$.  Define $w(s) = (1-s^2)^\alpha
y(2s^2-1)$, $s \in [0,1]$. Substituting $\beta = -1/2$ into equation
(\ref{jacobi ode 1}) it becomes
\begin{eqnarray}
  \left[\frac{w'(s)}{(1-s^2)^{\alpha-1}}\right]' +
  \frac{2(\alpha+n)(1+2n)w(s)}{(1-s^2)^{\alpha}} = 0,
\label{jacobi ode 2}
\end{eqnarray}
which is of the form
\begin{displaymath}
  [k(s)w'(s)]'+\phi(s)w(s) = 0
\end{displaymath}
with $k(s)$ and $\phi(s)$ positive, and $k(s)\phi(s)$ increasing, if
$\alpha$ and $n$ are positive. These are the necessary conditions for
the Sonine-P\"olya theorem (appendix \ref{sonine}), which states that
the local maxima of $|w(s)|$ will be decreasing. From its definition
$|w(s)|$ has a local maximum at $s = 0$, since $w(0)w''(0) < 0 $, and
a local minimum at $s = 1$, since $w(0) = 0$. $w(s)$ is continuous so
it is bounded by its local maxima, hence $|w(s)| \leq |w(0)|$, for $s
\in [0,1]$. In the original variables this is
\begin{eqnarray}
  \left(\frac{1-x}{2}\right)^\alpha
    \left|P^{(\alpha,-1/2)}_n(x)\right| & \leq &
    \left|P^{(\alpha,-1/2)}_n(-1)\right| \label{ode-1/2bound}.
\end{eqnarray}
Substituting in the values of $P^{(\alpha,-1/2)}_n(-1)$ and
$P^{(\alpha,-1/2)}_n(1)$ this becomes\footnote{The Pochhammer symbol
$(a)_n = \Gamma(a+n)/\Gamma(a)= a(a+1)\ldots(a+n-1)$.}
\begin{eqnarray}
  |\tilde{P}^{(\alpha,-1/2)}_n(x)| & \leq &
  \frac{(1/2)_n}{(\alpha+1)_n} \left(\frac{2}{1-x}\right)^\alpha ,
\end{eqnarray}
for $-1 \leq x \leq 1$. The right hand side is decreasing with $n$ if
$\alpha > -1/2$.  So for $n \geq 4$
\begin{eqnarray}
  |\tilde{P}^{(\alpha,-1/2)}_n(x)| & < &
   \frac{105/16}{(\alpha+1)_4}\left(\frac{2}{1-x}\right)^\alpha.
\end{eqnarray}
This is increasing with x so achieves its maximum at $x =
-(2\alpha+3)/(2\alpha+5)$. Thus
\begin{eqnarray}
  |\tilde{P}^{(\alpha,-1/2)}_n(x)| & \leq &
  \frac{105/16}{(\alpha+1)_4}
  \left(\frac{2\alpha+5}{2\alpha+4}\right)^\alpha .
\end{eqnarray}
For $\alpha \geq 1$ this is strictly bounded by
\begin{eqnarray}
  \frac{1}{(\alpha+1)(2\alpha+5)} & = &
  \left|\tilde{P}^{(\alpha,-1/2)}_1\left(-\frac{2\alpha+3}{2\alpha+5}\right)
  \right|,
\end{eqnarray}
and since it is decreasing and $x \leq -(2\alpha+3)(2\alpha+5)^{-1}$
\begin{eqnarray}
  |\tilde{P}^{(\alpha,-1/2)}_n(x)| & < &
  |\tilde{P}^{(\alpha,-1/2)}_1(x)|.
\end{eqnarray}
But $\tilde{P}^{(\alpha,-1/2)}_1(x)$ is negative on the range of $x$
so
\begin{eqnarray}
  \tilde{P}^{(\alpha,-1/2)}_n(x) & > & \tilde{P}^{(\alpha,-1/2)}_1(x)
  \,\,{\rule[0ex]{1.5ex}{1.5ex}}
\end{eqnarray}

\subsection{$CS^{d-1}$, $\beta = 0$}
\label{alpha:zero}
In ${\bf CS}^{d-1}$ the zonal spherical polynomials are
$P^{(\alpha,0)}_n(x)$, where $\alpha = d - 2$, and a similar theorem
exists.

{\bf Theorem 4} {\em
  \begin{equation}
    \label{inequalityb0}
    \tilde{P}^{(\alpha,0)}_n(x) > \tilde{P}^{(\alpha,0)}_1(x),
  \end{equation}
  for $-1 < x < -(\alpha+1)(\alpha+3)^{-1}$, $n > 1$ and $\alpha \geq
  2$, where $\tilde{P}^{(\alpha,0)}_n(x) =
  P^{(\alpha,0)}_n(x)/P^{(\alpha,0)}_n(1)$.  }

I begin by considering two special cases, $n=2$ and $n=3$.  The first
four polynomials are \widetext
\begin{eqnarray*}
  \tilde{P}^{(\alpha,0)}_0(x) & = & 1
  \\ 
  \tilde{P}^{(\alpha,0)}_1(x) & = & \frac{\alpha +
  (\alpha+2)x}{2(\alpha+1)}
  \\
  \tilde{P}^{(\alpha,0)}_2(x) & = &
  \frac{\alpha^2-\alpha-4+2\alpha(\alpha+3)x +
  (\alpha+3)(\alpha+4)x^2}{ 4(\alpha+1)(\alpha+2)}
  \\
  \tilde{P}^{(\alpha,0)}_3(x) & = & \frac{\alpha(\alpha^2-3\alpha-16)
  + 3(\alpha-3)(\alpha+2)(\alpha+4)x + 3\alpha(\alpha+4)(\alpha+5)x^2
  }{}\\&&
  \frac{\mbox{}+ (\alpha+4)(\alpha+5)(\alpha+6)x^3}{
  8(\alpha+1)(\alpha+2)(\alpha+3)}
\end{eqnarray*}
So
\begin{eqnarray}
  \label{p2-p1}
  \tilde{P}^{(\alpha,0)}_2(x)-\tilde{P}^{(\alpha,0)}_1(x) & = &
  -\frac{(\alpha+4)(1-x)(\alpha+1+(3+\alpha)x)}{
  4(\alpha+1)(\alpha+2)}
  \\  \label{p3-p1}
  \tilde{P}^{(\alpha,0)}_3(x)-\tilde{P}^{(\alpha,0)}_1(x) & = &
  -\frac{(\alpha+5)(1-x)[\alpha(3\alpha+8)+2(\alpha+4)(2\alpha+3)x
  +(\alpha+4)(\alpha+6)x^2}{8(\alpha+1)(\alpha+2)(\alpha+3)}
\end{eqnarray}
\narrowtext Equation (\ref{p2-p1}) is positive for $x <
-(\alpha+1)(\alpha+3)^{-1} $ (hence the range chosen for
(\ref{inequalityb0}).)  Equation (\ref{p3-p1}) is positive when the
quadratic factor
\begin{equation}
  \label{quadraticfactor2}
  \alpha(3\alpha+8)+2(\alpha+4)(2\alpha+3)x+(\alpha+4)(\alpha+6)x^2,
\end{equation}
is negative.  At $ x = - 1 $ equation (\ref{quadraticfactor2}) equals
$-4\alpha$, and at $ x = -(\alpha+1)/(\alpha+3)$ it equals
$-4(\alpha+2)(\alpha+6)(\alpha+3)^{-2}$ both of which are negative for
$\alpha > 0$. Therefore inequality (\ref{inequalityb0}) holds for $n =
2$ and $n = 3$.

The method in the previous appendix cannot be used unless $\beta =
\pm1/2$, since the differential equation has a singular point at $x =
-1$.  There is a simple method\footnote{Szeg\"o proves that for
polynomials $p(s)$ orthogonal with weight function $w(s)$, that if
$w(s)$ is non-decreasing then $[w(s)]^{1/2}|p(s)|$ is non-decreasing
also. The weight measure over which $P^{(\alpha,0)}$ are orthogonal is
$(1-x)^\alpha \mbox{dx}$. After changing variable to $x=2s^2-1$ the
new measure is proportional to $s^{(2\alpha+1)}\mbox{ds}$, which is
non-decreasing.} for the special case of $\beta = 0$, based on a
result from Szeg\"o (7.21.2) \cite{Szego}
\begin{equation}\label{temp2}
  \left[(1-x)/2\right]^{\alpha/2+1/4}\left|P^{(\alpha,0)}_n(x)\right|
    \leq 1,
\end{equation}
when $-1\leq x \leq 1$ and $\alpha \geq -1/2$.  Substituting
$P^{(\alpha,0)}_n(1) = (\alpha + 1)_n/n!$ into (\ref{temp2}) it
becomes
\begin{eqnarray}
  \left|\tilde{P}^{(\alpha,0)}_n(x)\right| & \leq &
    \frac{n!}{(\alpha+1)_n}
    \left(\frac{2}{1-x}\right)^{(\alpha/2+1/4)}.
\end{eqnarray}
For $\alpha > 0 $ the right hand side is decreasing with $n$, so for
$n \geq 4$
\begin{eqnarray}
  \left|\tilde{P}^{(\alpha,0)}_n(x)\right| & \leq &
    \frac{4!}{(\alpha+1)_4} \left(\frac{2}{1-x}
    \right)^{(\alpha/2+1/4)}.
\end{eqnarray} 
This is decreasing with $x$ so achieves its maximum at $x \leq
-(\alpha+1)(\alpha+3)^{-1}$. Thus
\begin{eqnarray}
  \left|\tilde{P}^{(\alpha,0)}_n(x)\right| & \leq &
    \frac{4!}{(\alpha+1)_4} \left(\frac{\alpha+2}{\alpha+3} \right)
    ^{(\alpha/2+1/4)}.
\end{eqnarray}
For $\alpha \geq 2$ this is strictly bounded by
\begin{eqnarray}
    \frac{1}{(\alpha+1)(\alpha+3)} &=&\left|
    \tilde{P}^{(\alpha,0)}_1\left(-\frac{\alpha+1}{\alpha+3}
    \right)\right|.
\end{eqnarray}
Since $|\tilde{P}^\alpha_1(x)|$ is monotonic increasing
\begin{eqnarray}
  \left|\tilde{P}^{(\alpha,0)}_n(x)\right| & < &
    \left|\tilde{P}^{(\alpha,0)}_1(x)\right|.
\end{eqnarray}
But $\tilde{P}^\alpha_1(x)$ is negative on the range so
\begin{displaymath}
  \tilde{P}^{(\alpha,0)}_n (x) > \tilde{P}^{(\alpha,0)}_1(x)
  \,\,{\rule[0ex]{1.5ex}{1.5ex}}
\end{displaymath}

\section{Sonine-P\"olya theorem}
\label{sonine}
This standard theorem is referred to in \cite[7.31.2]{Szego}.

{\bf Theorem 5} {\em Let $y(x)$ be a solution of the differential
equation
\begin{displaymath} 
  [k(x)y'(x)]'+\phi(x)y(x) = 0 \mbox{.}
\end{displaymath}
If $k(x)$ and $\phi(x)$ are positive, and $k(x)\phi(x)$ is increasing
(decreasing) and its derivative exists, then the local maxima of\/
$|y(x)|$ are decreasing (increasing).  }

Let
\begin{displaymath}
  f(x) = [y(x)]^2 + [k(x)y'(x)]^2[k(x)\phi(x)]^{-1}
\end{displaymath}
then $f(x) = [y(x)]^2$ if $y'(x)=0$, and
\begin{eqnarray*}
  f' & = & 2y' \left\{ y + \frac{[ky']'}{k\phi}
  -\frac{[k\phi]'y'}{2\phi^2} \right\} \\ & = & -\frac{y'^2 [k\phi ]
  '}{\phi^2}.
\end{eqnarray*}
So $\mbox{sgn} f'(x) =
-\mbox{sgn}[k(x)\phi(x)]'\,\,{\rule[0ex]{1.5ex}{1.5ex}}$

\section{An Example of Large Probability Violation}\label{dheg}
Consider the $2n$ vectors
\begin{eqnarray}
  ({\bf u}_i)_j = \frac{a\delta_{ij}-1}{\sqrt{a^2-2a+n}} \in {\bf
  H}_1,\\ ({\bf v}_i)_j = \frac{b\delta_{ij}+1}{\sqrt{b^2+2b+n}}\in
  {\bf H}_2,
\end{eqnarray}
where
\begin{eqnarray}
  a = \frac{1+\epsilon+\sqrt{(1+\epsilon)(1+\epsilon-n\epsilon)}}
  {\epsilon},\\ b =
  \frac{1-\epsilon+\sqrt{(1-\epsilon)(1-\epsilon+n\epsilon)}}
  {\epsilon}
\end{eqnarray}
and $\epsilon \leq 1/(n-1)$.  Then
\begin{equation}
  {{\bf u}_i}^\dagger{\bf u}_j = \delta_{ij}(1+\epsilon)-\epsilon
  \quad\mbox{and}\quad {{\bf v}_i}^\dagger{\bf v}_j =
  \delta_{ij}(1-\epsilon)+\epsilon.
\end{equation}
Define
\begin{equation}
  {\bf w}_i = ({\bf u}_i \oplus {\bf u}_i)/\sqrt{2} \in {\bf H}_1
  \oplus {\bf H}_2,
\end{equation}
so
\begin{equation}
  {\bf w}_i^\dagger{\bf w}_j = \delta_{ij}.
\end{equation}
Let the initial state be
\begin{equation}
  \psi = \frac{1}{\sqrt{n}}\sum_{i=1}^n{\bf w}_i \in {\bf H}_1 \oplus
  {\bf H}_2.
\end{equation}
Then use $\{{\bf w}_i{\bf w}_i^\dagger\}$ as a set of projectors to
get the history states
\begin{equation}
  \left\{{\bf w}_1/\sqrt{n},\ldots,{\bf w}_n/\sqrt{n}\right\}
\end{equation}
Then make a projection onto ${\bf H}_1$ and ${\bf H}_2$ to get the
history states
\begin{equation}
  \left\{{\bf u}_1/\sqrt{2n},\ldots,{\bf u}_n/\sqrt{2n},{\bf
      v}_1/\sqrt{2n},\ldots,{\bf v}_n/\sqrt{2n}\right\}.
\end{equation}
The decoherence matrix can be written
\begin{equation}
  \frac{1}{2n} \left(
    \begin{array}{cccccccc}
             1 & -\epsilon & \ldots & -\epsilon & 0 & \ldots & \ldots
             & 0 \\ -\epsilon & \ddots & \ddots & \vdots & \vdots &
             \ddots & \ddots & \vdots \\ \vdots & \ddots & \ddots &
             -\epsilon & \vdots & \ddots & \ddots & \vdots \\
             -\epsilon & \ldots & -\epsilon & 1 & 0 & \ldots & \ldots
             & 0 \\ 0 & \ldots & \ldots & 0 & 1 & \epsilon & \ldots &
             \epsilon \\ \vdots & \ddots & \ddots & \vdots & \epsilon
             & \ddots & \ddots & \vdots \\ \vdots & \ddots & \ddots &
             \vdots & \vdots & \ddots & \ddots & \epsilon \\ 0 &
             \ldots & \ldots & 0 & \epsilon & \ldots & \epsilon & 1
    \end{array}
  \right).
\end{equation}
The MPV for this set is $|-n(n-1)\epsilon/(2n)| = (n-1)\epsilon/2
\approx d\epsilon/4$. It is achieved by coarse-graining all the ${\bf
u}_i$'s (or ${\bf v}_i$'s) together.

\section{Quantum Zeno Effect}\label{zenoapp}
The Quantum Zeno effect is often discussed in the interpretation of
quantum mechanics, but has had no quantitative analysis in the
consistent histories formalism.

Consider a two dimensional Hilbert space.  Define the vectors
\begin{equation}
  {\bf u}^n_+ = \left(\begin{array}{c}\cos(n\epsilon)\\
\sin(n\epsilon)\end{array}\right), {\bf u}^n_- =
\left(\begin{array}{c}-\sin(n\epsilon)\\\cos(n\epsilon)\end{array}\right),
  \label{vectordef}
\end{equation}
and the projectors
\begin{equation}
  P^n_+ = {\bf u}^n_+ {{\bf u}^{n}_+}^\dagger, P^n_- = {\bf u}^n_-
  {{\bf u}^{n}_-}^\dagger.
  \label{projectordef}
\end{equation}
For any $n$, $P^n_+$ and $P^n_-$ are a complete set of projectors.
Consider the set of histories formed by using strings of these
projectors on the initial state ${\bf u}^0_+$.
\begin{equation}
  C_\alpha = P^n_{\alpha_n} \ldots P^1_{\alpha_1}.
\end{equation}
The histories $\alpha$ are string of $n$ pluses or minuses.

Define $|\alpha|$ to be the number of transitions from plus to minus
or vice versa in the string $\{\alpha_1,\ldots,\alpha_n,+\}$. Then
\begin{equation}
  C_\alpha {\bf u}^0_+ = {\bf
    u}^n_{\alpha_n}(-1)^{\lfloor\frac{|\alpha| +1}{2}\rfloor}
    \cos^{n-|\alpha|}\epsilon\sin^{|\alpha|}\epsilon,
\end{equation}
and there will be 
$n\choose|\alpha|$ identical histories states.  The non-zero
decoherence matrix elements are those with $|\alpha| = |\beta|
\bmod{2}$ and are
\begin{equation} \label{egdm}
  D_{\alpha\beta} = (-1)^{\lfloor\frac{|\alpha| +1}{2}\rfloor}
  (-1)^{\lfloor\frac{|\beta| +1}{2} \rfloor}
  \cos^{2n-|\alpha|-|\beta|}\epsilon\sin^{|\alpha|+|\beta|}\epsilon,
\end{equation}

Because of the simple form of (\ref{egdm}) all of the following
calculations can be done exactly, but for simplicity I shall let
$\epsilon = \theta/n$ and work to leading order in $1/n$.  The largest
probability violation for this decoherence matrix will be achieved by
coarse-graining together all the histories with a positive sign into
one history and all those with a negative sign into another.  Let $X$
denote the histories $|\alpha| = 0,3 \bmod 4$ and $Y$ the histories
$|\alpha| = 1,2 \bmod 4$. Then the probability violations for these
sets are,
\begin{eqnarray*}
  \sum_{\alpha\neq\beta\in X}D_{\alpha\beta} & = & 1/2 \cosh^2\theta +
  1/2 \cos\theta \cosh\theta
\\&&\mbox{}- 1/2 \sin\theta \sinh\theta - 1+ O(1/n)
\end{eqnarray*}
for $X$ and
\begin{eqnarray*}
  \sum_{\alpha\neq\beta\in Y}D_{\alpha\beta} &=& 1/2\cosh^2\theta -
  1/2\cos\theta\cosh\theta
  \\&&\mbox{}+  1/2\sin\theta\sinh\theta+ O(1/n)
\end{eqnarray*}
for $Y$.  The off-diagonal elements in the decoherence matrix
(\ref{egdm}) are all less than $\theta^2/n^2$ yet the MPV is order
$\exp{(2\theta)}$, so by choosing $n \gg \theta \gg 1$ the
off-diagonal elements can be made arbitrarily small whilst the MPV is
arbitrarily large. This proves the following theorem.

{\bf Theorem 6} {\em \label{proveuseless} For all Hilbert spaces of
dimension $\geq 2$, $\epsilon > 0$ and $x > 0 $ there exist finite
sets of histories such that
  \begin{equation}
    |D_{\alpha\beta}| \leq \epsilon, \quad \forall \alpha \neq \beta,
  \end{equation}
  and with $\mbox{MPV} > x$.  }

Now suppose the limit $n \to \infty$ is taken. Then all the elements
of the decoherence matrix (\ref{egdm}) are zero except for
$D_{\alpha\alpha} = 1$, $\alpha = \{+\cdots+\}$. A naive argument
would be to say that since all the off-diagonal elements are zero the
set is consistent, but this is false. The set is pathologically
inconsistent.

This shows that care must be taken with infinite sets of histories. It
is incorrect to take the limit of a set of histories and then apply
consistency criteria. Instead the order must be reversed and the limit
of the criteria taken. This does not always seem to have been
recognised in the literature. For instance Halliwell
\cite{Halliwell:fluctuations} states : ``In particular, it
[$|D_{\alpha\beta}| \leq (D_{\alpha\alpha}D_{\beta\beta})^{1/2}$]
implies that consistency is automatically satisfied if the system has
one history with $D_{\alpha\alpha} = 1$, and $D_{\beta\beta} = 0$ for
all other histories.'' He says this after a similar limit has been
taken, and I have shown above that this is not necessarily true.

The DHC trivially rejects this family of histories as grossly
inconsistent since
\begin{equation}
  \frac{D_{\alpha\beta}}{(D_{\alpha\alpha} D_{\beta\beta})^{1/2}} = 1,
  \quad \mbox{whenever $|\alpha| = |\beta| \bmod2$}.
\end{equation}


\begin{thebibliography}{10}

\bibitem{LL:QM} L.~D. Landau and E.~M. Lifschitz, {\em Quantum
Mechanics (Non-Relativistic Theory)}, Vol.~3 of {\em Course of
Theoretical Physics}, 3rd ed. (Pergamon Press, Oxford, 1977).

\bibitem{Espagnat} B. d'Espagnat, {\em Conceptual Foundations of
Quantum Mechanics} (Addison-Wesley, Reading, 1976).

\bibitem{GM:Hartle:classical} M. Gell-Mann and J.~B. Hartle,
Phys. Rev. D {\bf 47}, 3345 (1993).

\bibitem{Caldeira:Leggett} A.~O. Caldeira and A.J.Leggett, Physica A
{\bf 121}, 587 (1983).

\bibitem{Joos:Zeh} E. Joos and H.~Z. Zeh, Zeit. Phys. B {\bf 59}, 223
(1985).

\bibitem{Dowker:Halliwell} H.~F. Dowker and J.~J. Halliwell,
Phys. Rev. D {\bf 46}, 1580 (1992).

\bibitem{Pohle} H.-J. Pohle, Physica A {\bf 213}, 345 (1995).

\bibitem{Paz:brownian:motion} J.~P. Paz, in {\em The Physical Origins
of Time Asymmetry}, NATO, edited by J.~J. Halliwell,
J.~P. P\'erez-Mercader, and W.~H. Zurek (Cambridge University Press,
Cambridge, 1994).

\bibitem{Zurek:preferred:observables} W.~H. Zurek, in {\em The
Physical Origins of Time Asymmetry}, NATO, edited by J.~J. Halliwell,
J.~P. P\'erez-Mercader, and W.~H. Zurek (Cambridge University Press,
Cambridge, 1994).

\bibitem{Anastopoulos:Halliwell} C. Anastopoulos and J.~J. Halliwell,
gr-qc {\bf 9407039}, (1994).

\bibitem{Tegmark:Shapiro} M. Tegmark and H.~S. Shapiro, Phys. Rev. E
{\bf 50}, 2538 (1994).

\bibitem{Paz:Zurek:Classicality} J.~P. Paz and W.~H. Zurek,
Phys. Rev. D {\bf 48}, 2728 (1993).

\bibitem{Hartle:spacetime} J.~B. Hartle, Phys. Rev. D {\bf 44}, 3173
(1991).

\bibitem{Zurek:transition} W.~H. Zurek, Phys. Today {\bf 44}, 36
(1991).

\bibitem{Griffiths:1} R.~B. Griffiths, J. Stat. Phys. {\bf 36}, 219
(1984).

\bibitem{Omnes:1} R. Omn\`es, J. Stat. Phys. {\bf 53}, 893 (1988).

\bibitem{Isham:homog} C.~J. Isham, J. Math. Phys. {\bf 35}, 2157
(1996).

\bibitem{CEPI:GMH} M. Gell-Mann and J.~B. Hartle, in {\em Complexity
Entropy and the physics of Information}, Vol.~III of {\em SFI Studies
in the Science of Complexity}, Santa Fe Institute, edited by
W.~H. Zurek (Addison Wesley, Reading, 1990).

\bibitem{Omnes:interpretation} R. Omn\`es, Rev. Mod. Phys. {\bf 64},
339 (1992).

\bibitem{Omnes:2} R. Omn\`es, J. Stat. Phys. {\bf 53}, 933 (1988).

\bibitem{Omnes:3} R. Omn\`es, J. Stat. Phys. {\bf 53}, 957 (1988).

\bibitem{Omnes:4} R. Omn\`es, J. Stat. Phys. {\bf 57}, 357 (1989).

\bibitem{Omnes:5} R. Omn\`es, Ann. Phys. {\bf 201}, 354 (1990).

\bibitem{Omnes:truth} R. Omn\`es, J. Stat. Phys. {\bf 62}, 841 (1991).

\bibitem{Hartle:QM:Cosmology} J.~B. Hartle, in {\em Quantum Cosmology
and Baby Universes}, edited by S.  Coleman, J.~B. Hartle, T. Piran,
and S. Weinberg (World scientific, Singapore, 1989).

\bibitem{Dowker:Kent:approach} F. Dowker and A. Kent, J. Stat. Phys,
to appear, (1995) {\bf gr-qc/9412067}, (1994).

\bibitem{Dowker:Kent:Properties} F. Dowker and A. Kent,
Phys. Rev. Lett. {\bf 75}, 3038 (1995).

\bibitem{Halliwellconversation} J.~J. Halliwell, 1995, private
communication.

\bibitem{Shannon} C.~E. Shannon, Bell System Technical Journal {\bf
38}, 611 (1959).

\bibitem{Wyner} A. Wyner, Bell System Technical Journal {\bf 44}, 1061
(1965).

\bibitem{Mehta} M.~L. Mehta, {\em Random Matrices}, 2nd ed. (Academic
Press, London, 1991).

\bibitem{Kabatyanski:Levenshtein} G.~A. Kabatyanski and
V.~I. Levenshtein, Problems of Information Transmission {\bf 14}, 3
(1978).

\bibitem{Delsarte} P. Delsarte, J.~M. Goethal, and J.~J. Seidel,
Geometriae Dedicata {\bf 6}, 363 (1977).

\bibitem{Conway} J.~H. Conway and N.~J. Sloane, {\em Sphere Packings,
Lattices and Groups} (Springer-Verlag, New York, 1988).

\bibitem{Szego} G. Szeg\"o, {\em Orthogonal Polynomials} (American
Mathematical Society, New York, 1939).

\bibitem{Halliwell:fluctuations} J.~J. Halliwell, Phys. Rev. D {\bf
48}, 4785 (1993).

\end{thebibliography}
\end{document}